# A Review on Brain Mechanisms for Language Acquisition and Comprehension

**Kailash Nath Tripathi[1], Anand Bihari[2], Sudhakar Tripathi[3], Ravi Bhushan Mishra[4]**

[1]Merrut Institute of Technology, Merrut, [2]Vellore Institute of Technology, Vellore, [3]R. E. C. Ambedkar Nagar, U. P. [4]IIT (BHU) Varanasi

[1]kailash.tripathi@gmail.com, [2]anand.bihari@vit.ac.in, [3]p.stripathi@gmail.com, [4]ravibhushan@iitbhu.ac.in

**Abstract:**

This paper aims to bring the main viewpoint of language acquisition and language comprehension. In language acquisition, we have reviewed the different types of language acquisitions like first, second, sign and skill acquisition. The experimental techniques for neurolinguistic acquisition detection is also discussed. The findings of experiments for acquisition detection is also discussed, it includes the region of brain activated after acquisition. Findings shows that the different types of acquisition involve different regions of the brain. In language comprehension, native language comprehension and bilingual's comprehension has been considered. Comprehension involve different brain regions for different sentence or word comprehension depending upon their semantic and syntax. The different fMRI/EEG analysis techniques (statistical/ graph theoretical) are also discoursed in our review.Tools for neurolinguistic computations (pre-processing/computations/analysis) are alsodiscussed.

**1 Introduction**

The past several years has yielded an enormous research work in neuroscience investigating language acquisition, comprehension and production. Non-invasive, safe functional brain measurements have now been proven feasible for use with infants or adult for neural data acquisition. The neural signature of effect of learning at the phonetic level can be recognized at a amazingly high precision. Continuity in linguistic development, brain responses to even phonetic level stimuli can be observed with theoretical and clinical impact.

**2Language Acquisitions**

Human brain the command centre controls heart rhythm, memory and language to all human activities. Broca's area a small region in inferior frontal gyrus(IFG) necessary for production and coordination of language is found in left hemisphere in most of people. Wernicke's area the counter part of Broca's area in superior temporal gyrus(STG) performs language comprehension both written and spoken. The area of Broca's area is usually described as composed of the cytoarchitecturally defined area of Brodmann BA44, the pars opercularis and BA 45, and the pars triangularis.The cytoarchitecturally identified region BA 22 covers the latter two-thirds of the lateral convexity of the STG and is part of the Wernicke region.[1]

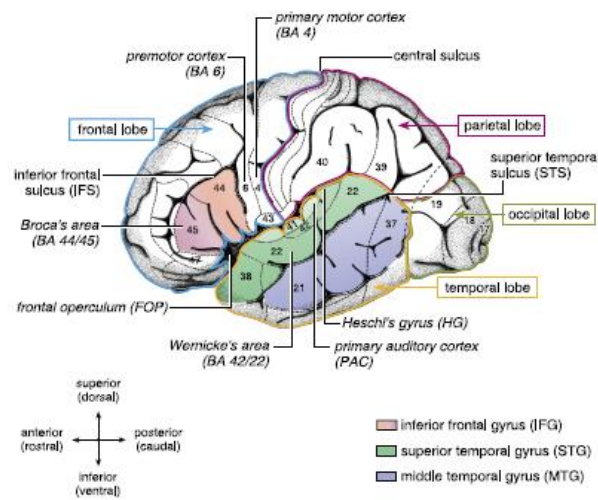

Figure 1: Langauge area in human brain comprises Broca's and Wernicke's Area

The acquisition of languages is one of the most important human traits and certainly it is the brain that undergoes the changes in development. Therefore the root of grammatical rules should be ascribed to an implicit process in the human brain. Linguists find speaking, signing and understanding language to be the key language skills, i.e. natural or inborn and biologically determined, while they find reading and writing to be secondary. In truth, acquisition of a native or first language (L1) through these primary faculties during the first years of life, whereas children learn their linguistic knowledge gradually. Speech in children progresses from babbling at around the age of 6-8 months, to the single-word stage at 10 to 12 months, and then to the two-word stage at around 2 years. There is a profound difference between linguistic factors between L1 and L2. An L2(Second Language) can be learned at any moment in life, although the L2 capacity is rarely comparable to that of L1 if it is acquired after the predicted 'sensitive period' from early childhood to puberty (È12 years of age). Numerous studies of functional magnetic resonance imaging(fMRI) and positron emission tomography (PET) have shown that auditory phonological processing is correlated with activation in the posterior superior temporal gyrus (STG) [Brodmann's region (BA) 22], while lexico-semantic processing is typically associated with activation in the left extra-Sylvian temporoparietal regions, including the angular ones [2].

In [3], Eric Lenneberg (1967) proposed that the acquisition of human language was an example of biologically limited learning, he stated that a child would have biological heritable component to learn language. He concluded that the process of acquiring langauge is profoundly ingrained and, species-specific, human biological property. Any language usually acquired during a crucial time beginning early in life and ending in puberty. He indicated that language could only be learned with difficulty or through a different learning method beyond this time.

A critical period is a time of maturation during which some of the key stimuli would have their peak impact on development or learning, resulting in normal actions adapted to the specific environment to which the organism was exposed. If the organism is not subjected to this phenomenon until after this period of time, the same phenomenon may have either a diminished effect, or may have no effect at all in extreme cases. Studies show a close association between age of language use and the ultimate degree of competency (PL) attained. However, exposure age does not affect all aspects of language leaning equally. Therefore, the crucial effects of the critical period seem to focus on phonology, morphology, syntax and not meaning processing [4].

The existence of critical, or at least a sensitive period for language acquisition in human being is explained by an evolutionary model suggested by J. R. Hurford in [5].

The acquisition of first language is one of the unexplained mysteries which surround us in our daily lives. A child learns language spontaneously, almost miraculously, as its learning of language progresses rapidly with an obvious pace and accuracy. Most children quickly learn language, giving the illusion that the process of acquiring first language is easy and straightforward. This is not the case, however, as children go through many stages of first language acquisition .The stages of language learning in children usually consist of: cooing, babbling, holophrastic stage, telegraphic speech and normal speech. The age of cooing is up-to 9 months till then children use phonemes from every language. At 9 month they start babbling in which they selectively use phonemes from their native language. At the age of 12 month they start using single words. When they are in holophrastic stage at around 18-24 months, they can combine words in two words stages.By the age of around 30 months they develop to the telegraphic stage where they can utter a clear phrase structure. As the children develop physically, so does their language skills as they internalize more complex systems by widening their vocabulary and their immediate surroundings. At the age of 5 years children reached up to normal developed speech.

There are three famous theories for first language acquisition: the behaviourist theory, the innatist theory and interactionists theory. Behaviourist theory[7] equated learning to a language all behaviour are acquired through interaction with environment and interactions are imitation, reinforcement, practice and habit formation. Children learn their first language by stimuli and children's responses are influenced by reinforcement.

The Innatist theory[8] believed that children are equipped with a device called the language acquisition device (LAD) and universal grammar (UG) which accounts for the swift mastery of language among children despite the extremely abstract nature of language. The Interactionists[9] believes that language is not a separate element of the mind as language reflects the information gained through children's physical contact with the world.

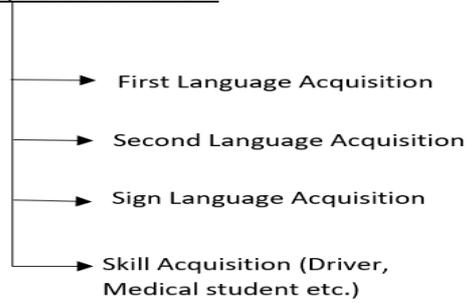

The acquisition of language in brain can be of four types: first language, second language, sign language and some additional skills which consists of some specialized/additional form of language. First language is the native language which is acquired by the infants naturally in the social environment. The learners have already learned at least one language in the case of second language acquisition and the prior experience in the first language that prove to be an advantage to them because they already have the idea of how language works. Second language learners often possess cognitive maturity and knowledge of metalinguistics that would be useful for them in solving problems when talking in second language [10]. Sign languages are used for communication in born deaf peoples. The learning and acquisition of sign language starts in later age well beyond the infants. To get the expertise in any skill, our brain need to acquire that skill by some special training or by practice, for example driver learns how to drive, medical student learns and acquire knowledge of her respective specialization and interpreter learns new foreign languages.

**2.1 First Language Acquisition**

Achieving language skills and language literacy in early childhood has been related to future reading performance and can influence academic achievement, mental wellbeing and potential job prospects [11], [12]. Neuroimaging studies also helped to develop a better understanding of the interaction between brain and language skills in white matter architecture. Studies show that young children rely on a large brain network to process languages, which becomes a more focused network with an increase in age [13].

Functional Near Infrared Imaging (fNIRS) studies[14] indicate that slow rhythmic modulations in the linguistic stream (< 8 Hz) mark syllable and word boundaries in the continuous linguistic stream, possibly helping children learn the words and structures of their language. The sensitivity of children to slow rhythmic modulations inherent in the linguistic stream facilitates the acquisition of language in childhood and the transition from speech to print language during the early years of reading[15], [16].

Neuroimaging research indicates that after hearing language, adult brain neural networks enter into a synchronized connection between the linguistic stream's various frequency modulations and the neural activity's endogenous rhythmic oscillations. Neuronal firing rates are known to oscillate at different frequency bands, including Delta (1-3 Hz), Theta (4-8 Hz) and Gamma (30-80 Hz). Slower frequencies match syllabic and word boundaries (Delta-Theta) and faster frequencies match individual phonemes[17], [18].

The results indicate that the right hemisphere may have an overall enhanced capacity to handle rhythmic response, while the left hemisphere may have a selected response to a preferred set of slow rhythmic modulations, which may be especially prominent for the brain system responsible for cross-modal language processing and reading[19].

Spatial language acquisition (SLA) consists of frame of reference (FoR) [20] which is acquired by individuals in their language (left/right, north/south). Languages vary widely in the availability and frequency of FoR terms [21]–[24]. For instance, English prefers egocentric terms ("left", "front") for describing small-scale table-top arrays, some other languages prefer geocentric terms like "north" or "uphill". In addition, a number of studies have shown a link between the dominant FoR in a particular language group and the availability of FoR representations in non-verbal cognitive tasks in community members[25], [26].

Authors in [27] conducted an experiment on children, and findings indicate the reliance on pre-existing language circuits for the acquisition of new native-phonology word types. This explains how after a few repetitions the children learn new vocabulary.

Authors in [28] conducted an fMRI experiment and demonstrated that delayed learning of a first language is correlated with changes in tissue concentration in the occipital cortex near the region that was found to display functional recruitment during language processing in persons with a late learning period. Such results indicate that a lack of familiarity with early language affects not only the functional but also the anatomical brain organization[29].

Authors in [30] found that the existence of feedback had a significant impact on the structure of the network used by learners to learn the properties of words in a natural language. A statistical learning system suggests that learners track distributional information in their environment and use that information to derive the structure and concepts they obtain about the sensory inputs. For example, in running speech, infants can segment words from an artificial language by monitoring the transitional probability of syllables.

**2.2 Second Language Acquisition**

Acquisition of their vocabulary is a crucial part of learning a new language. Morphology in the linguistic sense is the study of words, how they are created, and how they relate to other words in the same language. Study in [31] discussed the neural signature in initial phase of morphological rule based learning of a novel language (L2) in adults and suggests that even after a short exposure, adult language learners can acquire both novel words and novel morphological rules of L2.

Bilingualism studies have identified ways in which a second language's neural representation (L2) varies from that of the first language (L1) of an person[32]. In particular, there are many variations in activation between L2 and L1, both in degree and magnitude. L2s tend not only to display more activity within traditional language areas of the left hemisphere but also to enable more regions beyond the traditional language network. There are two prevailing hypotheses about why L2 neural signatures vary from L1 signatures. The first is that, during L2 learning, these variations reflect decreased neuroplasticity that occurs at a later age than learning with L1. L2 learning needs increased neural capital on this account due to maturational changes in neural plasticity within regions and pathways that enable first language learning [33]. The alternative hypothesis is that neural variations in L1 versus L2 are caused instead by the fact that the L2 of individuals is typically lower in ability than their L1. Therefore, the processing of L2 requires increased computational requirements and thus increased neural resources[34]. The experimental results indicate that ability and AoA describe different functional and structural networks within the bilingual brain, which we interpret as indicating distinct plasticity forms for age-dependent effects versus experience and/or skills.

Authors in [35] consider structural changes to brain areas believed to support language roles during learning of a foreign language. Experimental findings show that the volume of the hippocampus and the cortical thickness of the left middle frontal gyrus, inferior frontal gyrus and superior temporal gyrus increase for interpreters compared to controls. In interpreters with higher foreign language abilities, the right hippocampus and the left superior temporal gyrus were both structurally more maleable[36].

Study in[37] investigated how the age at which L2 was acquired influenced brain structures in bilingual people. This shows that AoA, language skills and current exposure rates are equally important in taking into account the systemic differences. Structural changes related to bilingualism and multilingualism have also been reported, bilinguals tend to have increased grey matter volume/density in Heschl's gyrus [38], the left caudate [39] and the left inferior parietal structure [40].

Authors in [41] explored the correlation between instructed second language acquisition (ISLA) skills and identified a clear connection between attitude towards language learning and second language skills. The analysis of language learning achievements in monozygotic and dizygotic twins [42] point to the possibility that having a positive attitude towards language learning and the language class is related to how well students do in ISLA independent of natural language abilities, teacher skill and L1-L2 relations.

In [43], authors examined the neural substrates of novel grammar learning in a group of healthy adults conducted an experiment and study based on fMRI that, in terms of functional connectivity, the involvement of the brain network during grammar acquisition is coupled with one's language learning ability.

**2.3 Sign Language Acquisition**

Children born deaf can not understand the languages spoken around them, and there is inadequate phonetic information provided by the visual signal of speech to facilitate spontaneous language acquisition. For many of these youngsters, language learning continues far beyond infancy after exposure to and immersion in a sign language at ages. Variation in the period of language acquisition in the adult brain influences language processing[44], [45]. In the classical language areas of the left hemisphere LH, fMRI studies of deaf native signers have find activation with a trend towards bilateral activation of the frontal and temporal lobes. These findings were observed using different tasks and triggers for distinct sign language namely American, British and Japanese[46]–[48].

The learning age is linearly and inversely related to activation rates in anterior language regions and positively related to activation rates in subsequent visual regions for linguistic tasks of American sign language (ASL) sentences, grammatical judgment and phonemic hand judgment [29].

## 2.4 Skill Acquisition

Authors in [49] addressed that the expression of this neuroplasticity depends on the age at which learning starts in several domains of skill acquisition. In studies aimed at determining the relationship between age of maturity and brain plasticity, the fact that most abilities are learned late in childhood or adulthood has proven to be a limit. According to [50], early sensory experiences tend to have the greatest capacity to improve neuronal circuitry in the early years of development, When the brain is in active building up phase. Neuroimaging studies of language development concentrate on the variations between simultaneous and concurrent bilinguals in brain structure and function, and whether bilingualism is accomplished later in life. It also discusses the idea of an optimal time in the production of languages and thus gives the relationship between the acquisition era and the ultimate results[51].

Santiago Ramon Y Cajal (Nobel prize winner) in 1894, proposed that mental activity might induced morphological changes in brain structure. Authors in[52] determined that the human brain structure expands and get renormalized during skill acquisition. It is known as the expansion-renormalization model, according to which neural processes related to learning always adopt a sequence of expansion, selection, and renormalisation. [53]. The model foretell an initial increase in the density of grey matter, theoretically representing the growth of neural capital such as neurons, synapses and glial cells, Accompanied by a selection process operating on this new tissue which results in a complete or partial return to the baseline overall volume after selection has been completed. To date, improvements in brain structure have been reported on different time scales, such as several months of juggling training, medical examination study, space navigation training, learning of foreign languages, etc.

For any language learning, the age of its acquisition matters a lot. The literature shows the importance of age for learning a language, early language acquisition improves the probability of being proficient in a language.For first language learning, social environment of infants also plays a significant role, age of learning, nature of input language and teaching strategy is also important. Second language acquisition becomes easier if it is learned in early age (before puberty)because during this period brain have more plasticity and it also have lot of idea about language learning which is experienced during first language learning. Vocabulary and grammar learning of second language is easier if it is done simultaneously or sequentially of L1 in early childhood. Sign language acquisition is done in later age than infants, as it is generally learned by born deaf children. Age of sign language learning also affects its proficiency. Skill development or expertise learning is also depending on age and the language proficiency before getting that skill. Learning at later stage can be improved by doing morphological learning.

**Table 2:** Review of Language acquisition in brain

| S.N. | Author | Task | Computation Method | Data Acquisition Method | Result |
|---|---|---|---|---|---|
| **Language Acquisition** | | | | | |
| 1 | P. K. Kuhl et. al., 2010[54] | Language and pre-reading in year two, third and fifth years. | Alpha, beta and gamma rhythms analysis. | EEG/ERPs/MEG/fMRI/NIRS | Early mastery of the phonetic units of language demands social learning. |
| 2 | R. I. Mayberry et. al., 2011[29] | American Sign Language, grammatical judgment and phonemic-hand judgment | t-statistics | fMRI data | The left lateralised activation pattern was observed |
| 3 | I. Kovelman et. al., 2012[14] | Language task and Rhythm Task | t-test analysis | functional Near Infrared (fNIRS) imaging | The right hemisphere overall displayed greater activation against the sluggish rhythmic stimulation, and the left hemisphere displayed greater activation compared to the quicker and slower frequencies. |
| 4 | J. Martensson et. al., 2012[35] | Three months of intense foreign language studies | t-test on cortical thickness | MRI | Structural changes in brain areas known for performing language roles during the learning of foreign languages. |
| 5 | S. Penicaud et. al., 2013[28] | American Sign Language (ASL) | voxel-based whole-brain correlational analysis. | fMRI | Not only the functional but also the structural structure of the brain is impaired by lack of early language experience. |
| 6 | Miao Wei et. al., 2015[37] | Language history questionnaire task | Cluster size, t-score. | MRI/fMRI/ PET | In the right parietal cortex, earlier second-language sensitivity is correlated with greater volumes. Consistently, as AoA decreased, the cortical region of the right superior |
| | | | | | parietal lobule increased. |
| 7 | E. Plante et. al., 2015[30] | Learners who spoke English were exposed to Norwegian sentences. | General Linear Model (GLM), Independent Component Analysis (ICA) | fMRI | The essence of the word input significantly affected the structure of the network used by the learners to learn the properties of words in a natural language. |
| 8 | I. A. Mendez et. al., 2015[41] | The parents and twin's questionnaires containing standard demographic questions and question assessing zygosity | Multiple regression (MR) analyses. | Cognition Based Statistical data | Lower language anxiety is related to higher abilities. Bilingualism and the starting age of directed second language learning (ISLA) often tend to be unrelated to language-learned proficiency. |
| 9 | E. S. Nichols et. al., 2016[32] | Picture-word matching task | TBSS (Tract-Based Spatial Statistics), Monte Carlo simulation. | fMRI and DTI | Within bilingual brain, Proficiency and AoA clarify different functional and structural networks. |
| 10 | A. Shusterman et. al., 2016[20] | Environment-based and Body-based Frame-of-References | t-tests | Cognition Based Statistical data | Findings suggest that it would be much more common to use the front and back axes to communicate about space than to use the world's languages. |
| 11 | J. A. Berken et. al., 2017[49] | Review on differences in brain structure and function between simultaneous and sequential bilinguals | Image based feature analysis of Grey matter density (GMD). | PET/fMRI/rsMRI | Simultaneous bilingual's brain function and structure appear to be most effectively organized. Sequential bilingual's ability for neuroplasticity change is apparently more constrained. |
| 12 | E. Wenger et. al., 2017[52] | Training of skill development task | Voxel-based morphometry (VBM) analysis | MRI | The provided model predicts an initial increase in the density of gray matter, theoretically reflecting the growth of neural capital such as neurons, synapses, and glial cells, followed by a selection mechanism operating on this new tissue leading to a complete or partial return to the baseline of the overall volume after the selection. |
| 13 | E. Partanen et. al., 2017[27] | Word form acquisition, associated with reading development | Event related field (ERF) waveform analysis. | MEG | The brains of the children seem more malevolent in learning novel word types than those of adults. A left-lateralized perisylvian network is often used by the developing brain to learn novel word types. |
| 14 | O. Kepinska et. al., 2017[43] | Grammar-learning task | Threshold-free cluster enhancement approach (TFCE), size of cluster, z-value | fMRI | With regard to functional communication, brain networks involvement during grammar acquisition is correlated with one's language learning abilities. |
| 15 | V. Havas et. al., 2017[31] | Early morphological learning of a novel language in adults | Analysis of Variance (ANOVA) on reaction times (RTs) | EEG | Adult language learners can acquire new words, as well as new morphological rules. |
| 16 | M. Walton et. al., 2018[11] | Assessments of Phonological Processing and Speeded Naming in children | Tract Based Spatial Statistics (TBSS). | DTI | Relationships seen in left ventricular pathways. Young children often rely on a large language processing network that gets more advanced with age. |

## 3 Language Comprehensions

Language processing refers to how human words are used to express thoughts and emotions. We as a neuroscience researcher are exploring how communications are processed and understood by the brain. Neuro-sensitive data based studies have shown that most of the language processing tasks are performed in the cerebral cortex. Most of the language role is handled in many different regions, and there are two well-identified regions considered essential to human language communication: the area of Wernicke and the area of Broca. The accurate fasciculus is the brain region between the Wernicke region and the Broca area which connects the two via bundles of nerve fibers. This part of the brain acts as a hub of transportation between the two areas mainly concerned with speech and communication.

Comprehension of sentences depends crucially on deciding the thematic relationship between noun phrases, i.e. defining who is doing what to whom. Study in [55] based on fMRI evaluated the relevant grammar and a key factor underlying the assessed output in the verbal working memory. Voxel-based gray matter morphometry showed that while the capacity of children to assign thematic roles in the left inferior temporal gyrus and the left inferior frontal gyrus in positively correlated with gray matter likelihood (GMP). The verbal work memory-related output in the left parietal operculum is positively associated with GMP extending into the posterior superior temporal gyrus. Those areas are known to be involved in dynamic sentence processing in a particular way. Results indicate a common GMP relationship in language-relevant brain regions and differential cognitive abilities that direct their interpretation of the sentence.

EEG mu rhythms recorded at fronto-central electrodes are commonly considered to be measures of human motor cortical activity as they are modulated when the participants perform an action, experience another's action or even imagine an action. Study in [56] recorded the modulation of mu rhythms in time frequency (TF), while participants interpreted the language of motion, abstract language and perceptive language. The findings indicate that mu repression is correlated with the language of practice rather than with abstract and perceptive language at fronto- locations. It also indicates that the activation takes place online through multiple words in the sentence, based on semantic integration.

During sentence processing region of the left upper temporal sulcus, inferior frontal gyrus and left basal ganglia show a systemic increase in brain activity as a function of constituent size, indicating their participation in computing syntactic and semantic structures. Experiments in [57] for non-spoken sign language on deaf participants show that the same network of language areas was found, while reading and sign language processing created similar effects of the linguistic structure in the basal ganglia, the effect of structure in all cortical language areas was greater for written language relative to sign language.

Based on evidence from neuroimaging, literature[58] reported both substantial overlap and unique linguistic cortical activation between comprehension of the sign language and observation of gestural behavior. In the upper / lower parietal lobe and the fusiform gyrus, overlaps in cortical activation are primarily observed. Authors in[59]found that American Sign Language (ASL) stimulated more strongly the left inferior frontal gyrus (IFG) and the middle superior temporal gyrus (STG) in deaf native signers than gestures expressing roughly the same material. Here Graph Theoretical Analysis(GTA) is used on the neural dependent cognition studies as an important complementary perspective to the activation research.

Study in [60] illustrates the semantic and grammatical processing of accented speech, both native and international. Closer analysis of listeners who did not understand the foreign accent correctly indicated that listeners who recognized the foreign accent displayed ERP responses for both grammatical and semantic errors. By comparison, listeners who did not correctly recognize the foreign accent gave no ERP responses to the foreign accented condition's grammatical errors, but displayed a late negativity to semantic errors.

Study in [61] indicates that mechanisms of the right hemisphere in the brain are essential to triggering elements of event information that breach the linguistic meaning. The brain stimulates components of event-knowledge that are semantically anomalous in context during learning.

In [62] authors propose that the prosodic information available during spoken language comprehension supports the generation of online predictions for upcoming words, and that comprehension of spoken language during serial visual presentation (SVP) reading, at least for quantifier sentences, may proceed more incrementally than understanding. The analysis demonstrates that the comprehension of spoken sentence continues fully incrementally, the results of truth meaning in both positive and negative quantifier sentences are alike. This also suggests that people use the spoken language more effectively than written SVP feedback to produce online predictions about coming words. During listening to natural speech, learning usually continues more incrementally than during an ERP experiment with N400 results during SVP hearing.

Authors in [63] note that, during late childhood and adolescence, the cortical depiction of language comprehension is added concentrated within the superior and middle temporal regions. Higher language ability are correlated with greater right hemispheric engagement during the listening of stories. Language comprehension is expressed more bilaterally than language output and a hemispheric dissociation with the development of the left hemispheric language, but comprehension of the bilateral or right hemispheric language is not uncommon even in healthy right handed subjects.

In [64], authors conducted an experiment and found that medial parietal lobe requires the production of referential words. Analysis of the experiment based on fMRI is done using a pairwise t test of total cluster activation which verified that each referential sub-condition was correlated with more activation than the non-reference condition.

The prefrontal brain regions historically associated with language comprehension are the Wernicke area and the Broca area.11 subjects of the Curtiss- Yamada Comprehensive Language Evaluation Receptive (CYCLE-R) are taken to perform voxel based lesion symptom mapping (VLSM) based analysis of functional neuroimaging data indicated that lesions to five left hemisphere brain regions affected performance on the CYCLE-R, including the posterior middle temporal gyrus and underlying white matter, the anterior superior temporal gyrus, the superior temporal sulcus and angular gyrus, mid frontal cortex in BA 46 and BA 47 of the inferior frontal gyrus. Analysis also suggested that the middle temporal gyrus may be more important for comprehension at the word level, while other regions may play a greater role at the level of the sentence.

**3.1 Bilingualism:** A large portion of the world's inhabitants is bilingual, and is flawlessly in over one language. A bilingual speaker routinely produces and understands without difficulty sentences which belong to two (or more) languages. Hence, knowing hoe two languages coexists in one brain with little disagreement or intrusion in both codes is a theoretical and applied question of great interest. There is ongoing debate about whether early and/or prolonged exposure to more than one language may lead to changes in patterns of brain activity during language processing.

Authors in [65] performed an experiment involving highly qualified bilingual Spanish / Catalan and Spanish monolinguals made grammatical and semantic decisions in Spanish while being tested for fMRI.Grammatical judgement showed increased activation in IFG (BA 45), fusiform gyrus (BA37), occipital lobe (BA 18) and in superior parietal lobe (SPL, BA 7). For monolingual group cortical activations were found in IFG (BA 45/46/9), SFG (BA 6), BA 8/32and BA 18/23/37). Study indicates bilinguals are attracting new areas of the brain. However, these different areas that depend on the learning age, language use, task circumstances, type of stimulus, cognitive / linguistic demands, and possibly the characteristics and relative similarities between the languages the bilingual speakers speak.

It has been shown that the two languages of Bilingual are simultaneously involved during listening, reading and speaking, even when only one language is specifically required. This parallel activation was shown to promote lexical access and to interfere in bilingual comprehension with the language processing. Research has shown that when bilinguals process visual words, they experience co-activation of language, and use inhibitory regulation to overcome non-target language competition. Authors in [66]suggest that the degree of language co-activation in bilingual spoken word comprehension is modulated by the amount of regular exposure to non-target language; and that bilinguals less affected by cross-language activation may also be more effective in suppressing non-linguistic task intervention.

Findings in [67] showed that language processing can be considered as the result of a network of brain regions interacting, rather than finding just a few brain areas to be involved in it. The experiment based on fMRI and its interpretation showed the activation of BA 44 and BA45 to be left lateralized in the three tasks (receptive semantic expressive paradigm), indicating roles in language phonology and semantic; however, their right homologous areas were also involved, which may be due to their involvement in executive function, attention or memory manipulation. On the contrary, BA 22 activation dominated at the right. The authors propose that right BA22's contribution to language acquisition is an integral part of a broader chain comprising left IFG, bilateral STG and lower parietal lobule. There are also studies that consider the right hemisphere as the seat for the transmission of phonology and semance.

EEG-based studies in[68]have taken on two tasks: the semantic decision-making task and the task of reading. The numerous experiment 1 wave maps indicate there was a frontal distribution of the disparity between literal and novel metaphoric sentences. For both studies, the amplitudes of late positive complex (LPC) for novel metaphoric sentences were decreased compared to those for anomalous sentences over parietal sites. While this effect was clearly lateralized in experiment 2, in experiment 1 it posed a wider parietal distribution.

Authors in [69] performed an experiment focused on repeated transcranial magnetic stimulation (rTMS) taking lexical decisions against basic tasks of judging. Findings provide evidence of an early motor cortex- TMS intervention protocol creates a lateralized left, task and meaning contextual improvement in response latencies, slowing down action-related word processing compared to faster abstract word reactions. The findings clearly suggest causal involvement of different modality circuits in language understanding, suggesting that cognitive phenomena of high order are based on simple biological mechanisms.

In [70], authors conducted a Near Infrared Spectroscopy (NIRS) experiment using the method of listening to English sentences with six separate speeches. The findings showed that Japanese subjects had understood speech with some of the characteristics of speech when amplitudes were expanded at certain frequency ranges. The NIRS measurement also revealed that the enhancement of high frequency amplitudes ranging from 7000 to 8500 Hz increased concentration of Oxy-Hb in most language areas (BA 45/44/22).

Study in[71] shows that the neural representation of sentences in two languages is normal. From a mapping built in English, the proposed model successfully predicted Portuguese sentences using brain positions and weights applied to neutrally plausible semant features (NPSF). The mapping between the neural activation patterns and

NPSF can be obtained in either language from any group of participants and yields positive activation prediction produced by a new sentence composed of new words.

Meta-language sentence prediction model: if the mapping between semiconductor and brain activation is similar across language, then a predictive model should be able to learn a mapping of semiconductor characterization and activation patterns in one language and predict the pattern of activation in another.

Study based on EEG experiment in[72] notes that there is a strong correlation between gamma band oscillations and semantic unification, while beta band oscillation has strong syntactic unification correlation.

Authors in [73] Introduce functional and anatomical connectivity to research a cognitive feature of interest subserving the network topology. Direct interactions between network nodes are defined here in a given network by analyzing functional time series of MRIs using the multivariate method of directional partial correlation (dPC). A region to region probabilistic fiber tracking on data from diffusion tensor imaging is performed to determine the most likely anatomical white matter fiber tracts that mediate the functional interactions for directly interacting pairs of nodes. The blended approach is extended to two stages of auditory comprehension: lowest understanding of speech and higher awareness of speech. Combining and applying interaction tracts of dPC and dorsal long and short reach, as well as commissural fibres.

The research in [74] suggested how the degree to which results relating oscillatory neural dynamics in the beta and gamma frequency ranges to the language comprehension of the sentence stage can be given a coherent description within a predictive coding system. They proposed that beta behavior represents both the active maintenance of the existing Neuro cognitive network (NCN) responsible for constructing and representing a sense of a sentence point, and the top-down dissemination of predictions based on that meaning to lower levels of the processing hierarchy.

The research in [75] revealed the creation of front-time resting state connectivity between adults and 5-year-olds by analyzing the association of intrinsic low-frequency BOLD oscillations in language-related regions. The results of a left and right IFG inter-hemisphere coupling in 5-year-olds and long-range correlation between IFG and pSTS in the left hemisphere in adults are consistent with previous low-frequency (LFF) analysis of fMRI evidence. Stronger long range communication in adults leads to a good limited left hemispheric language network growth trajectory. The findings support the notion that fronto-temporal functional connectivity is essential for the processing of syntactically complex sentences within the language network in the left hemisphere.

**Table 3:** Review of Language comprehension in brain

| S.N. | Author | Task | Computation Method | Data Acquisition Method | Result |
|---|---|---|---|---|---|
| **Language Comprehension** | | | | | |
| 1 | N. F. Dronkers et. al., 2004[76] | English sentence comprehension | t-test | fMRI/MRI | At word level, the middle temporal gyrus may be more important for comprehension, whereas at sentence level the other regions may play a greater role. |
| 2 | K. Lidzba et. al., 2011[63] | Beep stories (Language Comprehension) and language production (Vowel Identification) tasks. | Statistical analysis (t-tests). | fMRI | Only in the language comprehension test was verbal IQ correlated with lateralisation, with higher verbal IQ associated with more right-hemispheric participation. |
| 3 | A. Fengler et. al., 2015[55] | Standardized sentence comprehension test for determining the grammatical proficiency of participants | Voxel-based morphometry analysis(z-score, cluster size). | MRI | There is a clear correlation between the GMP of children in language-relevant brain regions and differential cognitive abilities which direct their understanding of the sentence. |
| 4 | A. G. Lewis et. al., 2015[72] | Semantic coherence in short stories, and other language comprehension tasks | beta and gamma oscillatory activity | EEG/MEG | Alternative proposal to link the beta and gamma oscillations for maintenance and prediction during language understanding. |
| 5 | P. Roman et. al., 2015[65] | Semantic infringement, loss of grammar and state of charge. | t-score | fMRI | Early bilingualism influences the brain and cognitive processes in the comprehension of sentences even in their native language; on the other hand, they indicate that brain over stimulation in bilinguals is not limited to a particular area. |
| 6 | I. Moreno et. Al., 2015[56] | Reading action language | Event-related potential (ERP) analysis and Time-frequency (TF) | EEG | Action language comprehension stimulates motor networks throughout the human brain. |
| | | | analysis | | |
| 7 | R. Metusalem et. Al., 2016[61] | Expected, Event-Related, Event-Unrelated words and comprehension question answers are used | Several statistical analyses were conducted on mean ERP voltage measures | EEG | Foster our understanding of the neural basis of event information activation and advance our understanding of how event awareness is activated in incremental understanding during creation of perceptions and elaborate inferences more generally. |
| 8 | D. Freunberger et. al., 2016[62] | N400 event-related potentials (ERP) | Linear Mixed Effects (LME) Models Using S4 Classes. | EEG | People use the spoken language more effectively than written SVP feedback to produce online predictions of coming words. |
| 9 | Y. Yang et. al., 2017[71] | English and Portuguese language reading | BOLD activation analysis | fMRI | Proven ability to predict meta-language through cultures, people and bilingual status. |
| 10 | S. Grey et. al., 2017[60] | Foreign-accented and native-accented speech | Mean ERP amplitudes | EEG/ERP | Provide novel insights into understanding the impact of listener familiarity and foreign-emphasized speaker status on language processing neural correlates. |
| 11 | L. Liu et.al., 2017[58] | Learning sign language by signers and Non-signers 'understanding of sign language. | Graph theoretical analysis (GTA) | fMRI | When observing sign language, the hearing signers and non-signers showed identical cortical activations. The frequently activated network was structured differently between the two classes, however. |
| 12 | C. Brodbeck et. al., 2017[77] | Visuo-spatial referential Domains | t-tests | MEG/EEG | Reports the medial parietal lobe participates in the production of referential words. |
| 13 | P. Chen et. al., 2017[66] | Word pairs consisting of an English-Korean inter-lingual homophone | ANOVAs with relatedness | Event-related potential (ERP) | The amount of regular exposure to the non-target language modulates the degree of language co-activation in bilingual spoken word comprehension. |
| 14 | N. Vukovic et. al., 2017[78] | Action words, abstract words and pseudo words | ANOVA, with the independent factors of Task | repetitive transcranial magnetic stimulation (rTMS) | Cortical motor regions play a vital role in understanding language. |
| 15 | K. Inada et. al., 2017 [70] | English speech task | Enhanced amplitudes | Near-infrared spectroscopy system (NIRS) | English discourses with enhanced amplitudes within a certain frequency range can affect brain function activation in the language processing area and contribute to a better understanding of English speaking. |
| 16 | A. Moreno et.al., 2018[57] | Sign language paradigm and written French stimuli | Z-score | MRI/fMRI | It suggests that the language network is systematically active in combinatorial language operations, comprising the left superior temporal sulcus, inferior frontal gyrus, and basal ganglia. |
| 17 | R. Alemi et. al., 2018[67] | Word Production (WP) task, Auditory Responsive Naming (ARN) paradigm, Visual Semantic Decision (VSD) paradigm | Group ICA | fMRI | The language function should be regarded as the result of a network of brain regions collaborating. |
| 18 | K. Rataj et. al., 2018[68] | Semantic decision and a reading task | t-test | EEG | The Late-Positive-Complex (LPC) pattern is modulated by both conventionality and task demand. |

## 4 Data Acquisition and Analysis Techniques

### 4.1 Data Acquisition

Over the last decade has shaped rapid developments in non-invasive practices that observe language processing in human brain. They include Electroencephalography (EEG)/ Event-related Potentials (ERPs), Magnetoencephalography (MEG), structural/resting-state Magnetic Resonance Imaging (rsMRI), functional Magnetic Resonance Imaging (fMRI), Near Infrared Spectroscopy (NIRS), Diffusion Tensor Imaging (DTI), Positron emission tomography (PET) etc.

**4.1.1 Magnetic resonance imaging (MRI)** can be paired with MEG and/or EEG, which offers static structural / anatomical brain images. Structural MRIs display structural variations over the lifetime of brain regions and they were recently used to predict second-language phonetic learning for adults. Magnetic resonance imaging (MRI) can be paired with MEG and/or EEG, which offers static structural / anatomical brain images. Structural

MRIs display structural variations over the lifetime of brain regions and have recently been used to predict phonetic learning of the second language of adults.[79]. In young children, structural MRI tests recognize the size of different brain structures and these tests have been shown to be linked to language skills later in childhood. When structural MRI images are superimposed on the physiological activity observed by MEG or EEG, it is possible to enhance the spatial localization of brain activity reported by those methods. [80].

**4.1.2 Functional magnetic resonance imaging (fMRI)** is a common tool for human neuroimaging, since it offers high spatial resolution maps of neural activity across the entire brain. [81]. The fMRI senses changes in bloodoxygenation that happen in the neural activation response. Neural effects occur in milliseconds; but the changes in bloodoxygenation they cause extend over many seconds, greatly restricting the temporal resolution of fMRI. fMRI learning let exact location of brain activity and some groundbreaking study illustrate remarkable similarities in the language-responsive structures in infants and adults. [82], [83].

**4.1.3 Electroencephalography (EEG)** is an electrophysiologic monitoring technique designed to capture brain electrical activity. This is characteristically non-intrusive, with the electrodes located around the scalp, but, as in electrocorticography, intrusive electrodes are sometimes used. EEG tests changes in voltage arising from ionic current inside brain neurons. In clinical contexts, EEG mentions the monitoring over a period of time of the normal electrical activity of the brain, as reported from multiple electrodes mounted on the scalp[84].

**4.1.4 Event-related Potentials (ERPs)** have been commonly used in infants and young children to study speech and language production. ERPs, a part of the EEG, represent electrical activity that is time-locked to present a particular sensory stimulus (for example, syllables or words) or a cognitive process (recognition within a sentence or phrase of a semantic violation)[85]. By placing sensors on a child's scalp, it is possible to quantify the behavior of neural networks firing in a synchronized and synchronous manner in open field environments, and to detect voltage shifts that occur as a result of cortical neural activity[86],[87].

**4.1.5 Magnetoencephalography (MEG)** Is another method for brain imaging which tracks exquisite temporal resolution of brain activity. The SQUID sensors positioned within the MEG helmet evaluate the minute magnetic fields associated with electrical currents generated by the brain while performing sensory, motor, or cognitive tasks. MEG facilitates the exact location of the neural currents accountable for magnetic field sources[88],[89] the use of modern head monitoring methods and MEG to illustrate phonetic recognition in newborns and infants in their first year of life..

**4.1.6 Near-Infrared Spectroscopy (NIRS)** Cerebral hemodynamic responses to neuronal activity are also measured, but light absorption sensitive to haemoglobin concentration is used to assess activation [90]. NIRS monitors increases in concentrations of blood oxy- and deoxy-haemoglobin in the brain, as well as increases in total blood volume in various areas of the cerebral cortex using near-infrared light. The NIRS system can assess activity in different brain regions by constantly monitoring the amount of haemoglobin in blood. In the first two years of life, studies have started to surface on children, testing infant responses to phonemes as well as longer periods of speech such as "motherese" and forward versus reversed sentences.

**4.1.7 Diffusion Tensor Imaging (DTI)** is an MRI-based neuroimaging technique that allows an estimation of the position, orientation, and anisotropy of the white matter tracts of the brain [91].

**4.1.8** Positron emission tomography (PET) tests pollutants from metabolically active chemicals injected into the bloodstream, which are radioactively labeled. The emission data is processed by a computer to generate multi-dimensional images of the distribution of the chemicals around the brain [92].

**4.2 Data Analysis**
Functional magnetic resonance imaging (fMRI) is a safe and non-invasive way of measuring brain function by using brain activity-related signal changes. The method has become an omnipresent instrument of fundamental, clinical, and cognitive neuroscience. This approach will calculate little changes in metabolism occurring in the active part of the brain. We analyze the fMRI data in order to identify the parts of the brain that are involved in a function, or to determine the changes that occur due to brain lesion in brain activities.

**4.2.1 Statistical Analysis Methods**
The efficiency of the fMR images is enhanced during the preprocessing stages. Thereafter, statistical analysis is attempted to establish which voxels the stimulus stimulates. Many of the fMRI studies are focused on the association between the hemodynamic response process and stimulation. Activation determines the changes in the images to local severity. These methods can be divided into two specific categories: univariate methods (methods for testing hypotheses), and multivariate methods (methods of exploration).

**4.2.1.1** The **univariate methods** seek to define which voxels, provided one signal model, can be defined as disabled. This allows response parameterisation and then model parameter estimation. The Generalized Linear Model (GLM) is method of univariate analysis [93].

**4.2.1.2** Multivariate approaches are often applicable to fMRI data analysis, which collects data from the sample, often with little prior knowledge of the conditions of experiment. They use certain structural properties, such as decorrelation, independence, similarity measures, which can discern characteristics of interest present in the data. Unlike the univariate methods conducting voxel-wise statistical analysis, multivariate methods provide statistical inference about the entire brain to explain spatial pattern brain responses.[94]. Multivariate method of analysis involves Concept Component Analysis (PCA), Independent Component Analysis (ICA), and Multi-Voxel Analysis of Patterns (MVPA). In MVPA feature Selection is made by approaches that pick the voxels that have more knowledge on the mental mission. There are many methods for the feature selection, including the t-test, f-score, ANOVA, the recursive feature evaluation metho[95].

## 5 Neuroimaging software tools

Software tools are used for analysis and visualization of neuro images to study the structure and function of the brain. Some of the popular neuroimaging software tools are: AFNI (Analysis of Functional NeuroImages), BrainSuite[96], CONN (Functional Connectivity Toolbox), EEGLAB, FreeSurfer, FSL and SPM (Statistical parametric mapping) etc.

**Table 1:** Tools for Neuro-data analysis

| S.No. | Tool Name | Availability | Input Data | Results |
|---|---|---|---|---|
| 1. | 3D Slicer (Slicer) [97] | Free and open source software | Image | Scientific visualization and image computing |
| 2. | Analysis of Functional NeuroImages (AFNI)[98] | Open source environment | functional MRI data | Mapping human brain activity |
| 3. | CONN[99] | Matlabbased cross-platform imaging software | fMRI and resting state MRI data | Computation, display and analysis |
| 4. | EEGNET [100] | MATLAB toolbox | Data from EEG, MEG, and other electrophysiological signals | ICA, time/frequency analysis, artefact rejection and several modes of data visualization. |
| 5. | FreeSurfer[101] | Brain imaging software package | MRI scan data | Functional brain mapping and facilitates the visualization of the functional regions of the highly folded cerebral cortex |
| 6. | Statistical parametric mapping (SPM)[102] | Matlab based toolbox | fMRI or PET | Statistical analysis |
| 7. | FMRIB Software Library (FSL)[103] | Freely available software library | functional, structural and diffusion MRI brain imaging data | Image and statistical analysis |
| 8. | Neuroimaging Informatics Tools and ResourcesClearinghouse (NITRC)[104] | Computational neuroscience tools and resources | MR, PET/SPECT, CT, EEG/MEG, optical imaging | Facilitating interactions between researchers and developers |

## 7 Conclusion

In this review paper we have shown how the brain behaves while language related tasks like language acquisition and language comprehension. We have found that most of the language related tasks are performed by Broca's and Wernicke's areas in the brain. In terms of Broadman Areas BA 22, BA44 and BA 45 are main ROIs for language related tasks. Literature also reveals that most of the other parts of the brain are also got activated while language comprehension depending upon the syntax and semantic of the sentences. IFG and STG also plays important role in sign language comprehension. Studies also shows that bilingual brains are more active than monolingual brains. We have also discussed about different data acquisition techniques for the study of brain behaviour. Different statistical analysis techniques are also discussed which is used for neuro data analysis.